\def\agt{\,%
  {\text{\raise-0.3ex\hbox{$\stackrel{>}{\sim}$}}\,}}
\def\alt{\,%
  {\text{\raise-0.3ex\hbox{$\stackrel{<}{\sim}$}}\,}}
\def\openone{1}
\def\sec#1{\par\noindent{\bf #1}}
\documentclass[draft]{book}
\usepackage{epsf}
\usepackage{nano2cmr}
\usepackage{amsfonts}
\advance\textheight by -1.2in
\begin{document}

\pnum{}
\ttitle{Refocusing of a qubit system coupled to an oscillator.}
\tauthor{{\em Leonid P.\ Pryadko} and Gregory Quiroz}

\ptitle{Refocusing of a qubit system coupled to an oscillator.}
\pauthor{{\em Leonid P.\ Pryadko} and  Gregory Quiroz}

\affil{Department of Physics \& Astronomy, University of California,
  Riverside, California, 92521, USA}

\begin{abstract}{Refocusing, or dynamical decoupling, is a coherent control
    technique where the internal dynamics of a quantum system is effectively
    averaged out by an application of specially designed driving fields.  The
    method has originated in nuclear magnetic resonance, but it was
    independently discovered in atomic physics as  a ``coherent destruction of
    tunneling''.  Present work deals with the analysis of the performance 
    of ``soft'' refocusing pulses and pulse sequences in protecting the
    coherence of a qubit system coupled to a quantum oscillator.  
  }
\end{abstract}

\begindc 

\index{Pryadko, L. P.}
\index{Quiroz, G.} 

\sec{Introduction and background.}  Quantum coherent control has found way
into 
many applications, including nuclear magnetic resonance (NMR), quantum
information processing (QIP), spintronics, atomic physics, etc.  The simplest
control technique is dynamical decoupling (DD), also known as refocusing, or
coherent destruction of tunneling.  The goal here is to preserve coherence by
averaging out the unwanted couplings.  This is achieved most readily by
running precisely designed sequences of specially designed pulses \cite{nmr}.

In a closed system this can be analyzed in terms of the average
Hamiltonian theory in the ``rotating frame'' defined by the
controlling fields \cite{avham}.  To leading order, the net evolution
over the refocusing period is indeed described by the Hamiltonian of
the system averaged over the controlled dynamics.  For an open system,
the dynamics associated with the bath degrees of freedom can be also
averaged out, as long as they are sufficiently slow.  This can be
understood by noticing that the driven evolution with period
$\tau=2\pi/\Omega$ shifts some of the system's spectral weight by the
Floquet harmonics, $\omega\to \omega+n\Omega$.  With the average
Hamiltonian for the closed system vanishing, the original spectral
weight at $n=0$ disappears altogether, and the direct transitions with
the bath degrees of freedom are also suppressed as long as $\Omega$
exceeds the bath cut-off frequency, $\Omega\agt\omega_c$
\cite{Kofman}.

For a {\em closed\/} system, the refocusing can be made more accurate
by designing higher-order sequences, where not only the average
Hamiltonian ($k=1$), but all the terms of order $k\le K$ in the Magnus
(cumulant) expansion of the evolution operator over the period $\tau$
are suppressed.  The corresponding simulation can be done efficiently
by constructing time-dependent perturbation theory on small clusters
\cite{SenguptaPRL}.  The quantum kinetics of the corresponding {\em
  open\/} system with order-$K$ DD, $K\le2$, was analyzed by one of
the authors using the non-Markovian master equation in the rotating
frame defined by the refocusing fields \cite{PryadkoPRB}.  This
involved a resummation of the series for the Laplace-transformed
resolvent of the master equation near each Floquet harmonic, with
subsequent summation of all harmonics.

The results of Ref.\ \cite{PryadkoPRB} can be summarized as follows.
With $K\ge1$ refocusing, there are no direct transitions, which allows
an additional expansion in powers of the small adiabaticity parameter,
$\omega_c/\Omega$.  In this situation the decoherence is dominated by
reactive processes (dephasing, or phase diffusion).  With $K=1$, the
bath correlators are modulated at frequency $\Omega$.  This reduces
the effective bath correlation time, and the phase diffusion rate is
suppressed by a factor $\propto\omega_c/\Omega$.  With $K=2$
refocusing, all 2nd-order terms involving instantaneous correlators of
the bath coupling are cancelled.  Generically, this leads to a
suppression of the dephasing rate by an additional factor
$\propto(\omega_c/\Omega)^2$, while in some cases (including
single-qubit refocusing) all terms of the expansion in powers of the
small adiabaticity parameter $(\omega_c/\Omega)$ disappear.  This
causes an \emph{exponential} suppression of the dephasing rate, so
that an excellent refocusing accuracy can be achieved with relatively
slow refocusing, $\Omega\agt\omega_c$.

The conclusions in Ref.\ \cite{PryadkoPRB} were based on the analysis of the
oscillator bath with a featureless spectral function, with the cut-off
frequency $\omega_c$ serving as the only scale describing the bath
correlations.  They do not apply in the presence of sharp spectral features
which appear if the controlled system is coupled to a local high-$Q$
oscillator.  On the other hand, the situation where the controlled system is
coupled to a local oscillator mode is quite common.  This situation is
realized in atomic physics, where the oscillator in question is the cavity
mode, while the continuous-wave (CW) excitation is used to suppress the
coupling. 
In several quantum computer designs, 
nearly-linear oscillator modes are inherently present (e.g., mutual
displacement in ion traps, or QCs based on electrons on helium).  Finally,
there are suggestions to include local high-$Q$ oscillators in the QC designs
to serve as ``quantum memory'' or ``quantum information bath'' \cite{QC-osc}.

In this work we consider refocusing of a qubit system where the spectral
function of the oscillator bath has a sharp resonance.  More specifically, we
include the resonant mode in the system Hamiltonian, and consider the quantum
kinetics of the resulting system in the presence of a featureless
low-frequency oscillator bath driven by the refocusing pulses applied to 
the qubits only.  Such a system can be analyzed with the help of the general
results \cite{PryadkoPRB}, as long as one is able to construct a $K=1$ or
$K=2$ refocusing sequence to decouple the oscillator and other degrees of
freedom.  To this end, and having in mind sequences of soft pulses, we
consider the analytical structure of the evolution operator for a closed
system of arbitrary complexity, where one of the qubits is driven by a single
one-dimensional $\pi$-pulse.  An analysis of any refocusing sequence is then
reduced to computing an ordered product of evolution operators for individual
pulses.  We illustrate the technique by analyzing the controlled dynamics of a
qubit coupled to an oscillator.  One of the constructed sequences provide
order $K=2$ qubit refocusing for any form of qubit--oscillator coupling, and
was also shown to provide an excellent decoupling in the presence of a thermal
bath.

\sec{Single $\pi$-pulse.}  Consider a qubit with generic couplings,
\begin{equation}
  H_S={\sigma_x}A_x+{\sigma_y}A_y+{\sigma_z}A_z+A_0, \label{eq:single-qubit}  
\end{equation}
where $\sigma_\mu$ are the qubit Pauli matrices and $A_\nu$, $\nu=0,x,y,z$ are
the operators describing the degrees of freedom of the rest of the system
which commute with $\sigma_\mu$, $[\sigma_\mu,A_\nu]=0$.  The qubit evolution
is driven by a one-dimensional pulse,
\begin{equation}
  H_C=\textstyle{1\over2}{\sigma_x}V_x(t), \quad 0<t<\tau_p,\label{eq:one-pulse}  
\end{equation}
where the field $V_x(t)$ defines the pulse shape.  
The evolution due to the pulse is dominant; the unitary evolution operator to
zeroth order in $H_S$ is simply 
\begin{equation}
  U_0(t)=e^{-i\sigma_x \phi(t)/2}, \quad \phi(t)\equiv \int_0^t
  dt'\,V_x(t').\label{eq:control-evolution}   
\end{equation}
When acting on the spin operators, this is just a rotation, e.g.,
$U_0(t)\sigma_yU_0^\dagger (t)=\sigma_y\cos \phi(t)+\sigma_z \sin
\phi(t)$.  Suppose $V_x(t)$ be symmetric, $V_x(\tau_p-t)=V_x(t)$,
$\pi$-pulse, $\phi(\tau_p)=\pi$, and it additionally satisfies the
first-order self-refocusing condition $s\equiv
\langle\sin\phi(t)\rangle_p=0$, where $\langle f(t)\rangle_p$ denotes
the time-average over pulse duration.  Then, the evolution operator
$X\equiv U_0(\tau_p)$
expanded to second order in $\tau_p H_S$ reads
  \begin{eqnarray}  \nonumber 
    \lefteqn{ X^{(2)}=  - i \sigma_x  -\tau_p ( A_x+ \sigma_x  A_0)
      + {i\over 2}\tau_p^2
      \{A_0 , A_x\}      }
    & &     \\ & & \hskip-0.2in\nonumber  
    + {i\over 2}\tau_p^2
    \sigma_x  (A_0^2+A_x^2)
    + \tau_p^2 \alpha  \bigl(A_y^2+A_z^2+ i  \sigma_x  [A_y ,  A_z]  \bigr) 
    \\ & &      \hskip-0.2in
    + \tau_p^2 \zeta \bigl( [A_0 , \sigma_y A_z-\sigma_z A_y] + i 
    \{A_x, \sigma_y A_y+\sigma_z A_z\}\bigr).\quad\strut
    \label{eq:pi-expansion}
  \end{eqnarray}
  Here $\alpha\equiv \langle \theta(t-t')\sin [\phi(t)-\phi(t')]\rangle_p$,
  $\zeta\equiv \langle\theta(t-t') \cos \phi(t')\rangle_p $ parametrize the
  evolution properties of the pulse at second order.  The values of the
  parameters computed for some pulse shapes are listed in
  Tab.~\ref{tab:params}.


\begin{table}[hc]
  \centering
  \begin{tabular}[c]{c|c|c|c|}
    pulse & $s\equiv \langle \sin\phi(t)\rangle_p$ & $\alpha/2$ & $\zeta$ \\ 
    \hline 
    $\pi\delta(t-\tau_p/2)$ & 0 & 0 & 0.25 \\ 
    $G_{001}$ & 0.0148978 & 0.00735798 & 0.249979\\
    $G_{010}$ & 0.148979  & 0.0653938 & 0.247905 \\ 
    $S_1$ \cite{SenguptaPRL}& 0 & 0.0332661 & 0.238227 \\
    $S_2$ \cite{SenguptaPRL}& 0 & 0.0250328 & 0.241377 \\ 
    $Q_1$ \cite{SenguptaPRL}& 0 & 0 & 0.239889 \\     
    $Q_2$ \cite{SenguptaPRL}& 0 & 0 & 0.242205 
  \end{tabular}
  \caption{Parameters of several common pulse shapes.  The first line
    represents  the ``hard'' $\delta$-function pulse, $G_{001}$ denotes the
    Gaussian pulse with the width $0.01\tau_p$, while $S_n$ and $Q_n$ denote
    the 1st and 2nd-order self-refocusing pulses from Ref.~\cite{SenguptaPRL}.}
  \label{tab:params}
\end{table}
\sec{Common pulse sequences.}
Transforming Eq.~(\ref{eq:pi-expansion}) appropriately, we can now easily
compute the result of application of any pulse sequence.  In particular, the
$\pi$-pulse $\overline X$ applied along the $-x$ direction can be obtained from
$(-X)$ with a substitution $\alpha\to-\alpha$.  As a result, e.g., the expansion
of the evolution opeator for the one-dimensional sequence $\overline X X$ can be
written as
$$
\overline X X=
\openone -2 i \tau_p (
     A_0 + \sigma_x  A_x) -2 \tau_p^2    (A_0 +\sigma_x A_x)^2
+\mathcal{O}(\tau_p^3),
$$
or it  can be re-exponentiated as evolution with the effective Hamiltonian
\begin{equation} 
  H_{\overline X X}=
  A_0 +\sigma_x  A_x+\mathcal{O}(\tau_p^2). \label{eq:1d-symmetric}
\end{equation}
The corresponding calculation with the usual finite-width pulses
(e.g., Gaussian) where $s\neq0$ produces a correction to the effective
Hamiltonian already in the leading order,
$$
\delta H_{\overline X X}=s (\sigma_z A_y-\sigma_y A_z)
+\mathcal{O}(s^2\tau_p).  
$$
This can be corrected by constructing a longer sequence, e.g., $\overline X X
X\overline X$.  Returning to pulses with $s=0$, we list the expansions
computed for several two-dimensional sequences:
\begin{eqnarray}
  \nonumber 
  \label{eq:four-pulse}
  \lefteqn{ H_{X\overline Y XY}=A_0 +\tau_p\Bigl( {i\alpha \over
      2}[A_z,A_y]-{i\over2} [A_0,\sigma_x A_x-\sigma_y A_y]} & & \\
  & & 
  -{\alpha\over
    2}\sigma_y(A_x^2+A_y^2) 
  +{1+4\zeta\over4}\sigma_z \{A_x,A_y\}
  \Bigr)+\mathcal{O}(\tau_p^2),
  \hskip0.2in\strut\\
  \lefteqn{ H_{YX\overline Y XX\overline Y XY}=A_0 -{\alpha\tau_p\over 2}
    \Bigl(\sigma_y (A_x^2+A_z^2)+i[A_y,A_z]\Bigr),}
  \label{eq:8p}
  \\
  \lefteqn{ H_{\overline Y\overline X Y\overline XX\overline Y XY}=A_0
    +\mathcal{O}(\tau_p^2).} 
  \label{eq:8a}
\end{eqnarray}
\sec{Atom in a cavity example.}  Consider an atom placed in a lossless cavity
with a single resonant mode.  The resonance part of the Hamiltonian can be
written in the form (\ref{eq:single-qubit}), where $A_x=g (b+b^\dagger)$,
$A_y=ig (b-b^\dagger)$, $A_z=0$, $A_0=\Delta\, b^\dagger b$, $g$ is a coupling
constant, and $\Delta$ is
the cavity frequency bias.  Contrary to the conclusions of Ref.~\cite{thor},
such a coupling cannot be suppressed with any one-dimensional pulse sequence.

On the other hand, the two-dimensional four-pulse
sequence~(\ref{eq:four-pulse}) provides a leading-order refocusing of the
coupling.  The subleading-order correction is present (the order of the
sequence is $K=1$), and it is not particularly small even for 2nd-order
self-refocusing pulses with $s=\alpha=0$. The eight-pulse sequences
(\ref{eq:8p}) and (\ref{eq:8a}) have equal accuracy with 2nd-order pulses but
the 2nd-order accuracy of the latter sequence is also retained with 1st-order
pulses.

After tracing out the oscillator degrees of freedom, we can apply the results
of Ref.~\cite{PryadkoPRB} and expect the two 2nd-order sequences to provide an
excellent refocusing accuracy even in an open system, as long as the refocusing
rate is sufficiently high.  We confirmed this expectation by a numerical
simulation, where the bath was modeled as a classical correlated random field.

\sec{Conclusions.}  The main result of this work is the expansion
(\ref{eq:pi-expansion}) and the classification of the corresponding parameters
in Tab.~\ref{tab:params}.  This allows an explicit computation of the error
operators associated with refocusing in systems of arbitrary complexity.  We
illustrated the approach for several sequences applied to a qubit coupled to
an oscillator.  The 8-pulse sequence~(\ref{eq:8a}) provides 2nd-order
refocusing for any form of the coupling between the qubit and the oscillator.

\sec{Acknowledgements.} This research was supported in part by the NSF
grant No.\ 0622242 (LP) and the Dean's Undergraduate Research fellowship
(GQ).

\end{document}